\documentclass[prc,showpacs,preprintnumbers,amsmath,amssymb]{revtex4-1}
\usepackage{bm}
\begin{document}
\title{Properties of Skyrme force as a residual interaction in beyond mean-field theories}
\author{Mitsuru Tohyama}
\affiliation{Faculty of Medicine, Kyorin University, Mitaka, Tokyo
  181-8611, Japan \email{tohyama@ks.kyorin-u.ac.jp}}
\begin{abstract}
In an effort to find an effective interaction which can consistently be used for both the mean-field part
and the residual part in beyond mean-field theories, 
properties of the Skyrme interactions as a residual interaction 
are investigated. The time-dependent density-matrix theory (TDDM) is used as a beyond mean-field theory and
the ground states of $^{16}$O and $^{40}$Ca are calculated using the five standard parametrizations of the Skyrme interaction which differs in density and momentum dependence. 
It is found that the Skyrme interaction which has strong density dependence and weak momentum dependence induces substantial ground-state correlations comparable to the results
of other theoretical calculations.

\end{abstract}
%\keyword{Extended RPA theory, Isovector dipole giant resonance}
\maketitle
\section{Introduction}
The time-dependent Hartree-Fock theory (TDHF) is the basis of the mean-field theories 
such as the Hartree-Fock theory (HF) and the random-phase approximation (RPA): A stationary solution of the TDHF equation gives the HF ground state and 
RPA can be formulated as the small amplitude limit of the TDHF equation. Since the introduction of the Skyrme interactions which well describe ground state properties of nuclei in HF \cite{vaut},
the Skyrme HF and self-consistent HF+RPA approaches have extensively been used as standard methods to study nuclear structure problem \cite{RS}.
Extensive TDHF simulations have also been performed for heavy-ion reactions \cite{Davi,sekiza}. Most experimental data, however, suggest 
that beyond-mean field theories which include two-body correlation effects are required for a realistic description of nuclear structure and reactions.
The time-dependent density matrix (TDDM) approach \cite{toh2020} is one of such beyond mean-field methods derived by truncating a coupled chain of
the equations of motion for reduced density matrices known as 
the Bogoliubov-Born-Green-Kirkwood-Yvon (BBGKY) hierarchy \cite{Bonitz}. The TDDM equations determine the time evolution of both one-body and two-body density matrices.
A stationary solution of the TDDM equations gives a correlated ground state, the small amplitude limit of the TDDM equations corresponds to
an extended RPA and also the TDDM equations describe two-body dissipations in heavy-ion collisions. 
In most TDDM simulations the Skyrme interactions are used for the mean-field part but the interaction used for the beyond mean-field level usually takes 
the form of a simplified $\delta$-function to facilitate numerical calculations of two-body matrix elements \cite{toh2020,assie,wen}.
In principle the interaction used for the residual channels should be consistent with that used for the mean-field part.  
Recently Barton et al. \cite{barton} have performed such consistent TDDM calculations. They
studied the ground-state correlations in light nuclei using a fully unrestricted three-dimensional
implementation of TDDM and the same Skryme interactions for both the mean-field part and the residual channels.
To avoid difficulties in dealing with a density-dependent force as a residual interaction \cite{barton,gamba}, 
they used the Skyrme parametrizations SV \cite{beiner} and SHZ2 \cite{satula} that do not have a density dependent term ($t_3$ term).
They found quite small ground-state correlations in contrast to the TDDM results obtained with the simple force \cite{toh2020}
and also to the results of other theoretical calculations \cite{adachi,taka,utsuno}.
The reason why the Skyrme interactions without the density dependence induce small ground-state correlations has not been analyzed in their work, however.
Obviously more studies on the residual-channel properties are needed for a wide range of the Skyrme interactions
to obtain an effective interaction to be used in the consistent TDDM simulations.
In this paper the TDDM calculations are performed for the ground states of $^{16}$O and $^{40}$Ca using the Skyrme parametrizations SII, SIII, SIV, SV and SVI 
\cite{vaut,beiner} which have different density and momentum dependence, 
and their properties in the residual channels are investigated to clarify why some Skyrme parametrizations
show small correlation effects and also to find a candidate effective interaction
which can consistently be used for both the mean-field potential and the residual channels.
The paper is organized as follows. The TDDM equations are given in Sect. II. The reduction of a three-body interaction to an effective density-dependent
two-body interaction is explained in Sect. III. The results for $^{16}$O and $^{40}$Ca are presented in Sect. IV and Sect. V is devoted to summary.

\section{Formulation}
\subsection{TDDM equations}
\subsection{Time-dependent density-matrix theory and truncation schemes}
The TDDM equations and truncations schemes are explained in Ref. \cite{toh2020} but are presented below for completeness. 
In TDDM it is assumed that the total Hamiltonian $H$ consists of a kinetic energy term and a two-body interaction. The TDDM equations consist of
the coupled equations of motion for the one-body density matrix (the occupation matrix) $n_{\alpha\alpha'}$
and the correlated part of the two-body density matrix $C_{\alpha\beta\alpha'\beta'}$ ($C_2$).
These matrices are defined as
\begin{eqnarray}
n_{\alpha\alpha'}(t)&=&\langle\Phi(t)|a^+_{\alpha'} a_\alpha|\Phi(t)\rangle,
\\
C_{\alpha\beta\alpha'\beta'}(t)&=&\rho_{\alpha\beta\alpha'\beta'}(t)
%\nonumber \\
-(n_{\alpha\alpha'}(t)n_{\beta\beta'}(t)-n_{\alpha\beta'}(t)n_{\beta\alpha'}(t))
,
\label{rho2}
\end{eqnarray}
where $|\Phi(t)\rangle$ is the time-dependent total wavefunction
$|\Phi(t)\rangle=\exp[-iHt] |\Phi(t=0)\rangle$ and $\rho_{\alpha\beta\alpha'\beta'}$ is the two-body density matrix (
$\rho_{\alpha\beta\alpha'\beta'}(t)=\langle\Phi(t)|a^+_{\alpha'}a^+_{\beta'}a_{\beta}a_{\alpha}|\Phi(t)\rangle$).
Units $\hbar=1$ are used hereafter. In the following it is assumed that the single-particle states are time-independent.
The equations of motion for $n_{\alpha\alpha'}$ and $C_{\alpha\beta\alpha'\beta'}$ are derived from
\begin{eqnarray}
i \dot{n}_{\alpha\alpha'}&=&\langle\Phi(t)|[a^+_{\alpha'}a_{\alpha},H]|\Phi(t)\rangle
\label{n0}
\\
i\dot{\rho}_{\alpha\beta\alpha'\beta'}&=&\langle\Phi(t)|[a^+_{\alpha'}a^+_{\beta'}
 a_{\beta}a_{\alpha},H]|\Phi(t)\rangle,
\label{C0} 
\end{eqnarray}
by evaluating the commutation relations. They are written as
\begin{eqnarray}
i \dot{n}_{\alpha\alpha'}&=&
\sum_{\lambda}(\epsilon_{\alpha\lambda}{n}_{\lambda\alpha'}-{n}_{\alpha\lambda}\epsilon_{\lambda\alpha'})
\nonumber \\
&+&\sum_{\lambda_1\lambda_2\lambda_3}
[\langle\alpha\lambda_1|v|\lambda_2\lambda_3\rangle C_{\lambda_2\lambda_3\alpha'\lambda_1}
%\nonumber \\
-C_{\alpha\lambda_1\lambda_2\lambda_3}\langle\lambda_2\lambda_3|v|\alpha'\lambda_1\rangle],
\label{n}
\end{eqnarray}
\begin{eqnarray}
i\dot{C}_{\alpha\beta\alpha'\beta'}&=&
\sum_{\lambda}(\epsilon_{\alpha\lambda}{C}_{\lambda\beta\alpha'\beta'}
+\epsilon_{\beta\lambda}{C}_{\alpha\lambda\alpha'\beta'}
%\nonumber \\
-\epsilon_{\lambda\alpha'}{C}_{\alpha\beta\lambda\beta'}
-\epsilon_{\lambda\beta'}{C}_{\alpha\beta\alpha'\lambda})
\nonumber \\
&+&B_{\alpha\beta\alpha'\beta'}+P_{\alpha\beta\alpha'\beta'}+H_{\alpha\beta\alpha'\beta'}+T_{\alpha\beta\alpha'\beta'},
\nonumber \\
\label{N3C2}
\end{eqnarray}
where $\epsilon_{\alpha\alpha'}$ is the single-particle energy including the mean field and is given by
\begin{eqnarray}
\epsilon_{\alpha\alpha'}=\langle \alpha|t|\alpha'\rangle
+\sum_{\lambda_1\lambda_2}
\langle\alpha\lambda_1|v|\alpha'\lambda_2\rangle_A 
n_{\lambda_2\lambda_1}
\label{hf}.
\end{eqnarray}
Here $t$ is the kinetic energy, $v$ is the two-body interaction and the subscript $A$ means that the corresponding matrix is antisymmetrized. 
The term $B_{\alpha\beta\alpha'\beta'}$ in Eq. (\ref{N3C2}) consists of only the occupation matrices and describes  2 particle (p) -- 2 hole (h)
and 2h--2p
excitations, while $P_{\alpha\beta\alpha'\beta'}$ and $H_{\alpha\beta\alpha'\beta'}$
contain $C_2$ and express
p--p (and h--h) and p--h
correlations to infinite order, respectively \cite{WC,GT}.
The $T_{\alpha\beta\alpha'\beta'}$ term gives the coupling to the three-body correlation matrix ($C_3$)
\begin{eqnarray}
T_{\alpha\beta\alpha'\beta'}&=&\sum_{\lambda_1\lambda_2\lambda_3}
[\langle\alpha\lambda_1|v|\lambda_2\lambda_3\rangle C_{\lambda_2\lambda_3\beta\alpha'\lambda_1\beta'}
%\nonumber \\
+\langle\lambda_1\beta|v|\lambda_2\lambda_3\rangle C_{\lambda_2\lambda_3\alpha\alpha'\lambda_1\beta'}
\nonumber \\
&-&\langle\lambda_1\lambda_2|v|\alpha'\lambda_3\rangle C_{\alpha\lambda_3\beta\lambda_1\lambda_2\beta'}
%\nonumber \\
-\langle\lambda_1\lambda_2|v|\lambda_3\beta'\rangle C_{\alpha\lambda_3\beta\lambda_1\lambda_2\alpha'}],
\label{T-term}
\end{eqnarray}
where $C_{\alpha\beta\gamma\alpha'\beta'\gamma'}$ is
given by
\begin{eqnarray}
C_{\alpha\beta\gamma\alpha'\beta'\gamma'}=\langle\Phi(t)|a^+_{\alpha'}a^+_{\beta'}a^+_{\gamma'}a_{\gamma}a_{\beta}a_{\alpha}|\Phi(t)\rangle
%\nonumber \\
-{\mathcal AS}(n_{\alpha\alpha'}\rho_{\beta\gamma\beta'\gamma'}),
\end{eqnarray}
Here, ${\mathcal AS}$ is an operator which properly symmetrize and anti-symmetrizes $n_{\alpha\alpha'}\rho_{\beta\gamma\beta'\gamma'}$ under the exchange of the single-particle indices
such as $\alpha\leftrightarrow\beta$ and $\alpha'\leftrightarrow\beta'$.
Approximations for $C_3$ are needed to close the equations of motion within $n_{\alpha\alpha'}$ and $C_2$.
In the truncation scheme of Refs. \cite{WC,GT} $C_3$ is simply omitted.
In this work the following truncation scheme  is used where $C_3$ are given by
\begin{eqnarray}
C_{\rm p_1p_2h_1p_3p_4h_2}&=&\sum_{\rm h}C_{\rm hh_1p_3p_4}C_{\rm p_1p_2h_2h},
\label{purt1}\\
C_{\rm p_1h_1h_2p_2h_3h_4}&=&\sum_{\rm p}C_{\rm h_1h_2p_2p}C_{\rm p_1ph_3h_4}.
\label{purt2}
\end{eqnarray}
Here p and h refer to particle and hole states, respectively.
These 2p1h-2p1h and 1p2h-1p2h components of $C_3$ are the leading-order terms in perturbative expansion of $C_3$ using
the Coupled-Cluster-Doubles (CCD)-like ground state wavefunction \cite{ts14}.
These components of $C_3$ in Eq. (\ref{N3C2}) can be interpreted as self-energy contributions to the 2p--2h and 2h--2p components of $C_2$ and 
play a role in preventing overshoot of 2p--2h excitations when the residual interaction is strong \cite{ts14}.
The trace relation between the one-body and two-body density matrices $n_{\alpha\alpha'}=\sum_\lambda \rho_{\alpha\lambda\alpha'\lambda}/(N-1)$ is not conserved when any approximation is made for $C_3$.
It was pointed out \cite{ts14} that the fulfillment of the trace relation is drastically improved by using Eqs. (\ref{purt1}) and (\ref{purt2}). 
The conservation of the total energy and total particle number is not affected by the truncation schemes 
for $C_3$ as long as its symmetry and anti-symmetry properties under the exchange of single-particle indices is respected.

\subsection{Adiabatic method}

The ground state in TDDM is given as a stationary solution of the time-dependent equations 
(Eqs. (\ref{n}) and (\ref{N3C2})) which satisfies $\dot{n}_{\alpha\alpha'}=0$ and $\dot{C}_2=0$. Two methods have been employed to obtain the stationary solution. One is the adiabatic method
: Eqs. (\ref{n}) and (\ref{N3C2}) are solved by starting from the HF configuration and gradually increasing the strength of the residual interaction such as  
$v({\bm r}-{\bm r'})\times t/T$. This method is based on the Gell-Mann-Low theorem \cite{gell}
and has often been used to obtain approximate ground states with various time-dependent functionals \cite{toh2020,assie,wen,barton}. 
To suppress oscillating components which come from the mixing
of excited states, $T$ must be chosen to be much larger than the longest period in the system considered.
The other method is a usual iterative gradient method which is useful to obtain a rigorously stationary solution. Since it involves matrix inversion, the application of the gradient method is limited to small systems: 
The gradient method has been employed to obtain the ground states of the oxygen and calcium isotopes \cite{toh07,toh18} using several single-particle states around the Fermi level.

\section{Skyrme parametrizations and effective density-dependent two-body interaction}
The density-independent standard parametrization of the Skyme force is used 
to avoid complications in the treatment of a density-dependent effective interaction in the residual channels \cite{barton,gamba}. 
It consists of two-body and three-body parts.
The two-body part of the Skyrme force is given by \cite{vaut}
\begin{eqnarray}
v_2&=&t_0(1+x_0 P^\sigma)\delta^3({\bm r})
%\nonumber \\
+\frac{1}{2}t_1(k'^2\delta^3({\bm r})
%\nonumber \\
+\delta^3({\bm r})k^2)
%\nonumber \\
+t_2{\bm k}'\delta^3({\bm r})\cdot{\bm k},
\label{t0x0}
\end{eqnarray}
where $P^\sigma$ is the spin exchange operator, ${\bm r}={\bm r}_1-{\bm r}_2$, ${\bm k}=(\nabla_1-\nabla_2)/2i$ acts on the right and ${\bm k}'=-(\nabla_1-\nabla_2)/2i$ on the left. 
The two-body spin-orbit force is given by
\begin{eqnarray}
v_{LS}=i W({\bm \sigma}_1+{\bm \sigma}_2)\cdot {\bm k}'\times \delta^3({\bm r}){\bm k},
\label{ls}
\end{eqnarray}
where $\sigma$ is the spin operator.
The three-body part of the Skyrme interactions is written as $v_3=t_3\delta^3(\bm{r}_1-\bm{r}_2)(\bm{r}_2-\bm{r}_3)$. 
The Skyrme parametrizations SII, SIII, SIV, SV and SVI \cite{vaut,beiner} are given in Table \ref{tab1} in the increasing order of $t_3$.
These parameter sets have different contributions of the $t_3$ term and the momentum-dependent $t_1$ and $t_2$ terms to the repulsive part of the effective interaction.
\begin{table}
\caption{Skyrme force parameters}
\begin{center}
\begin{tabular}{c|ccccc} \hline
  & SV & SIV  & SII & SIII & SVI  \\ \hline
  $t_3$ (MeVfm$^6)$ & 0 & 5000& 9331.1 & 14000 & 17000  \\ 
$t_0$ (MeVfm$^3)$ &-1248.29 & -1205.6 & -1169.9 & -1128.75 & -1101.81 \\
$t_1$ (MeVfm$^5)$ & 970.56 &765 & 586.6 & 395 & 271.67\\
$t_2$ (MeVfm$^5)$ & 107.22 & 35 & -27.1 & -95 & -138.33\\
$x_0$  & -0.17 & 0.05 & 0.34 & 0.45 & 0.583\\
$W$ (MeVfm$^4)$ & 150 & 150 & 105 & 120 & 115\\ \hline
\end{tabular}
\label{tab1}
\end{center}
\end{table}

Since it is difficult to deal with a three-body force in the TDDM approach, it is necessary to reduce the $t_3$ term to a density dependent two-body interaction to be used in the residual channel.
For this purpose 
the ground-state expectation value of
the three-body part $V_3$ of the total Hamiltonian is considered, where $V_3$ is given by
\begin{eqnarray}
V_3=\frac{1}{6}\sum_{\alpha\beta\gamma\alpha'\beta'\gamma'}\langle\alpha\beta\gamma|v_3|\alpha'\beta'\gamma'\rangle a^+_\alpha a^+_\beta a^+_\gamma a_{\gamma'}a_{\beta'}a_{\alpha'}.
\nonumber \\
\label{t3}
\end{eqnarray}
The ground state expectation value of Eq. (\ref{t3}) is decomposed into the three parts according to the decomposition of 
the three-body density matrix $\rho_{\alpha\beta\gamma\alpha'\beta'\gamma'}$
\begin{eqnarray}
\rho_{\alpha\beta\gamma\alpha'\beta'\gamma'}&=&\langle \Phi_0|a^+_{\alpha'} a^+_{\beta'} a^+_{\gamma'} a_{\gamma}a_{\beta}a_{\alpha}|\Phi_0\rangle
\nonumber \\
&=&{\cal AS}(n_{\alpha\alpha'}n_{\beta\beta'}n_{\gamma\gamma'}+n_{\alpha\alpha'}C_{\beta\gamma\beta'\gamma'})
%\nonumber \\
+C_{\alpha\beta\gamma\alpha'\beta'\gamma'}.
\end{eqnarray}
In the following it is assumed that the occupation matrix is diagonal, that is, $n_{\alpha\alpha'}=n_\alpha\delta_{\alpha\alpha'}$. 
Then  
\begin{eqnarray}
\langle\Phi_0|V_3|\Phi_0\rangle&=&\frac{1}{6}\sum_{\alpha\beta\gamma\alpha'\beta'\gamma'} \langle\alpha\beta\gamma|v_3|\alpha'\beta'\gamma'\rangle 
\nonumber \\
&\times&[{\cal AS}(\delta_{\alpha\alpha'}\delta_{\beta\beta'}\delta_{\gamma\gamma'} n_\alpha n_\beta n_\gamma
%\nonumber \\
+n_{\alpha}\delta_{\alpha\alpha'}C_{\beta\beta'\gamma\gamma'})
%\nonumber \\
+C_{\alpha\beta\gamma\alpha'\beta'\gamma'}].
\label{V3}
\end{eqnarray}
A similar decomposition can be made in the equations of motion for $n_{\alpha\alpha'}$ and $C_2$ (Eqs. (\ref{n0}) and (\ref{C0})).
For even-even nuclei the first term in the parentheses consisting of $n_\alpha$'s
is expressed by the following density-dependent two-body interaction as discussed in Ref. \cite{vaut}
\begin{eqnarray}
\frac{1}{6}t_3(1+P^\sigma)\rho\delta^3(\bm{r}),
\label{t31}
\end{eqnarray}
where $\rho$ is the nuclear density. This interaction is inappropriate as a residual two-body interaction in TDDM because exchange effects with the single-particle states in $\rho$ 
cannot be taken \cite{gamba}.
The second term in the parentheses of Eq. (\ref{V3}) which involves $C_2$ tells us how the residual two-body interaction should look like. The second term 
becomes
\begin{eqnarray}
&\frac{1}{2}&\sum_{\alpha\beta\gamma\alpha'\beta'}(\langle\alpha\beta\gamma|v_3|\alpha'\beta'\gamma\rangle n_\gamma C_{\alpha'\beta'\alpha\beta}
%\nonumber \\
-\langle\alpha\beta\gamma|v_3|\gamma\alpha'\beta'\rangle n_\gamma C_{\alpha'\beta'\alpha\beta}
\nonumber \\
&-&\langle\alpha\beta\gamma|v_3|\alpha'\gamma\beta'\rangle n_\gamma C_{\alpha'\beta'\alpha\beta}).
\label{vnc}
\end{eqnarray}
The sum over $\gamma$ in the first term in the above equation gives the total density $\rho({\bm r})=\sum_{\alpha}n_\alpha|\phi_\alpha({\bm r})|^2$,
where $\phi_\alpha({\bm r})$ is the single-particle wavefunction.  The sums in the second and third terms in Eq. (\ref{vnc}) depend on 
the charge and spin state of the single-particle state $\alpha$ or $\beta$. The assumptions that a single-particle state is labeled by a spin coordinate
$\sigma$ neglecting spin-orbit coupling and that the nuclear density is independent of spin state, that is, 
$\rho_\uparrow(r)=\rho_\downarrow(r)=\rho(r)/2$ for each charge state, give the following density dependent two-body residual interaction $v_2(\rho)\delta^3(\bm{r}_1-\bm{r}_2)$
to be used in the TDDM calculations, 
where
\begin{eqnarray}
v_2(\rho)=\left\{
\begin{array}{ll}
t_3\rho_n & \mbox{for proton-proton} \\
t_3\rho_p & \mbox{for neutron-neutron} \\
\frac{1}{2}t_3 \rho & \mbox{for proton-neutron}.
\label{t32}
\end{array}
\right.
\end{eqnarray}
Here, $\rho_p$ and $\rho_n$ stand for the proton and neutron densities, respectively. 
In the following calculations for $^{16}$O and $^{40}$Ca it is further assumed that $\rho_p\approx\rho_n\approx\rho/2$.
The last $C_3$ term in Eq. (\ref{V3}) evaluated with Eqs. (\ref{purt1}) and (\ref{purt2}) is found quite small and can safely be neglected also in the residual channels.

\section{Results}
\subsection{$^{16}${\rm O}}
For the calculation of $n_{\alpha}$ and $C_2$ in $^{16}$O, the minimal single-particle space is used, which consists of the proton and neutron $1p_{1/2},~1p_{3/2}$ and $1d_{5/2}$ states
along the lines of most previous TDDM calculations \cite{toh2020,assie,wen,barton}.
The two-body interaction consisting of Eqs. (\ref{t0x0}) and (\ref{t32}) is used as the residual interaction
in the TDDM equations. The contribution of the spin-orbit force Eq. (\ref{ls}) was found small and is omitted from the residual interaction: 
A TDDM calculation including the spin-orbit force showed that 
the change in the occupation probabilities is less than $2\times10^{-3}$. 
The Coulomb interaction between protons is also neglected. The TDDM equations are solved using the adiabatic method:
The residual interaction is multiplied by $t/T$ with $T=2400$ fm/c.
To facilitate the TDDM simulations, the matrix elements of the residual interaction calculated at $t=0$ with the HF single-particle wavefunctions are used throughout a time evolution.

\subsubsection{Comparison of TDDM and EDA}
 First the results in TDDM are compared with those in exact diagonalization approach (EDA) to confirm the validity of the TDDM approach.
The occupation probabilities calculated in TDDM for $^{16}$O using SVI which induces the largest ground-state correlations are shown in Table \ref{tab2}. 
The results in EDA (in the parentheses) are obtained using the same single-particle states and residual interaction as those used in TDDM. 
The results in TDDM agree well with the EDA results.
\begin{table}
\caption{HF single-particle energies $\epsilon_\alpha$ and the occupation probabilities 
$n_{\alpha}$ calculated in TDDM for $^{16}$O using SVI. The results in EDA are given in the parentheses.}
\begin{center}
\begin{tabular}{c rr rr} \hline
 &\multicolumn{2}{c}{$\epsilon_\alpha$ [MeV]}&\multicolumn{2}{c}{$n_{\alpha}$}\\ \hline 
orbit & proton & neutron  & proton & neutron  \\ \hline
$1p_{3/2}$ & -15.8 & -19.3 & 0.920(0.927)& 0.920(0.926)  \\
$1p_{1/2}$ & -10.2 & -13.6 & 0.846(0.854) & 0.844(0.852)  \\
$1d_{5/2}$ & -4.3 & -7.6 & 0.104(0.097) & 0.105(0.099)  \\\hline
\end{tabular}
\label{tab2}
\end{center}
\end{table}
\begin{table}
\caption{Occupation probabilities of the proton single-particle states and the correlation energy calculated 
in TDDM for $^{16}$O. The value in the parentheses indicates the EDA result.}
\begin{center}
\begin{tabular}{c|ccccc} \hline
  & SV & SIV  & SII & SIII & SVI  \\ \hline
  $1p_{3/2}$ & 0.985 & 0.990&0.984 & 0.961 & 0.920  \\ 
$1p_{1/2}$ &0.978 & 0.988 & 0.982 & 0.932& 0.846 \\
$1d_{5/2}$ & 0.018 & 0.010 &0.017 & 0.048 & 0.104\\
$E_{\rm cor}$ (MeV) & -7.1 & -3.4 & -6.2 & -14.1& -28.0 (-27.2)\\ \hline
\end{tabular}
\label{tab3}
\end{center}
\end{table}

\subsubsection{Results for SII, SIII, SIV, SV and SVI}
The proton occupation probabilities calculated in TDDDM using SII, SIII, SIV, SV and SVI are summarized in Table \ref{tab3}.
The neutron occupation probabilities are similar to the proton values as shown in Table \ref{tab2} and are not given here.
The parameter sets
SII, SIV and SV induce small ground-state correlations, SIII does moderately and SVI strongly.
The small ground-state correlations induced by SV are consistent with the results of Ref. \cite{barton}.
The occupation probabilities calculated with SVI are comparable to the results of shell-model calculations \cite{utsuno},
which give 0.920, 0.820 and 0.071 to the $1p_{3/2},~1p_{1/2}$ and $1d_{5/2}$ states, respectively.
Let us try to explain why SV, SIV and SII induce small ground-state correlations.
The 2p--2h matrix elements of the residual interaction are essential to induce ground-state correlations and they involve
larger relative momenta than the matrix elements used for the mean-field potential because particle states have larger momenta than hole states.
The parameter sets SV, SIV and SII have weaker density dependence and stronger momentum dependence than SIII and SVI.
Therefore, in the 2p-2h matrix elements of 
SV, SIV and SII a large cancellation can occur between the attractive $t_0$ term and the repulsive $t_1$ and $t_2$ terms.
This is not the case in the matrix elements for the mean-field potential which involve lower relative momenta.
A TDDM calculation using SV and neglecting the $t_1$ and $t_2$ terms in the residual interaction was performed to confirm such a cancellation, 
and strong ground-state correlations were found:
The occupation probabilities obtained for the proton $1p_{3/2},~1p_{1/2}$ and $1d_{5/2}$ states
are 0.840, 0.799 and 0.174, respectively.

The correlation energy $E_{\rm cor}$ is also shown in Table \ref{tab2}: $E_{\rm cor}$ is given by $C_2$ as 
\begin{eqnarray}
E_{\rm cor}=\frac{1}{2}\sum_{\alpha\beta\alpha'\beta'}\langle\alpha\beta|v|\alpha'\beta'\rangle C_{\alpha'\beta'\alpha\beta}.
\end{eqnarray}
The correlation energy in EDA for SVI is also given in the parentheses.
The correlation energy decreases with increasing ground-state correlations ($|E_{\rm cor}|$ increases).
On the other hand 
the mean-field energy $E_{\rm MF}$ given by 
\begin{eqnarray}
E_{\rm MF}&=&\sum_{\alpha}\langle \alpha|t|\alpha\rangle n_\alpha
%\nonumber \\
+\frac{1}{2}\sum_{\alpha\beta\alpha'\beta'}\langle\alpha\beta|v|\alpha\beta\rangle_A n_\alpha n_\beta
\end{eqnarray}
increases with increasing ground-state correlations due to
the partial filling of the particle states and the depletion in the occupation of the hole states, which somewhat compensates the decrease 
in $E_{\rm cor}$. In the case of SVI $E_{\rm cor}=-28.0 $MeV and the increase in $E_{\rm MF}$ is 17.5 MeV. 
As a consequence the total energy is decreased by 10.5 MeV, which is 8.3 \% of the total energy in HF. 
The 10.5 MeV decrease in the total energy is also comparable to the shell model result of 9.5 MeV \cite{utsuno}.

In the past, SIII has been used as a residual interaction in variational shell-model calculations \cite{otsuka} and it was shown that SIII 
successfully describes ground-state properties of neutron rich light nuclei. This somewhat contradicts the present study where 
SIII only induces moderate ground-state correlations which are much smaller than the results of the shell-model calculations \cite{utsuno}.
The reason for this discrepancy is in the fact that the $t_3$ term used as the residual interaction in Ref. \cite{otsuka} is not Eq. (\ref{t32}) but Eq. (\ref{t31}). 
A TDDM calculation was performed using Eq. (\ref{t31}) and it was found that it induces strong ground-state correlations comparable to the SVI results: the occupation probabilities of 
the proton $1p_{1/2},~1p_{3/2}$ and $1d_{5/2}$ states are 0.898, 0.905 and 0.100, and $E_{\rm cor}$ is $-26.5$ MeV. 
These values are also close to the results obtained  
from the simple interaction consisting of only the $t_0$ and $t_3$ terms of SIII which has been used in previous TDDM simulations \cite{toh2020}:
The occupation probabilities of 
the proton $1p_{1/2},~1p_{3/2}$ and $1d_{5/2}$ states obtained from the simple interaction are 0.880, 0.904 and 0.104, and $E_{\rm cor}$ is $-28.4$ MeV. 
The factor 1/6 in Eq. (\ref{t31}) significantly reduces the contribution of the repulsive $t_3$ term.
As mentioned above, exchange properties to be fulfilled as a two-body residual interaction are not properly respected in  Eq. (\ref{t31}).

\subsection{$^{40}${\rm Ca}}
The ground-state correlations in $^{40}$Ca are also studied using the $2s_{1/2},~1d_{3/2},~1d_{5/2}$ and $1f_{7/2}$ states for both protons and neutrons
and following the approach used in Ref.\cite{toh18}: The simplified TDDM equations which include
only the 2p--2h and 2h--2p components of $C_2$ and neglect $C_3$ are solved using the gradient method. 
It has been shown for $^{16}$O \cite{toh15} that this approximation (the omission of $C_3$ and other components of $C_2$) well
reproduces $n_\alpha$'s in EDA. It has also been pointed out \cite{toh18} that the inclusion of 
the ph--ph, 2p--2p and 2h--2h components of $C_2$, which are  approximated by
$C_{\rm p_1h_1p_2h_2}=\sum_{\rm ph}C_{\rm p_1ph_1h}C_{\rm hh_2p_2p}$,
$C_{\rm p_1p_2p_3p_4}=\sum_{\rm hh'}C_{\rm p_1p_2hh'}C_{\rm hh'p_3p_4}/2$ and 
$C_{\rm h_1h_2h_3h_4}=\sum_{\rm pp'}C_{\rm h_1h_2pp'}C_{\rm pp'h_3h_4}/2$, respectively, improves the result of $E_{\rm cor}$.
The occupation probabilities calculated in TDDM for $^{40}$Ca using SII, SIII, SIV, SV and SVI are summarized in Table \ref{tab4}.
The correlation energy which is calculated using all the components of $C_2$ mentioned above is also given in Table \ref{tab4}. 
The results for $^{40}$Ca are similar to those for $^{16}$O: SII, SIV and SV induce small ground-state correlations, SIII does moderately and SVI strongly.
In the case of SIII and SVI
the depletion of the occupation probability of the $2s_{1/2}$ state is much smaller than that of the $1d_{3/2}$ state although their single-particle energies are similar.
This is presumably due to the fact that the $2s_{1/2}$ state has higher momentum components than the $1d_{3/2}$ state, enhancing cancellation of the $t_0$ and $t_3$ terms
by the momentum dependent $t_1$ and $t_2$ terms. In the case of SVI
$E_{\rm cor}$ is $-56.5$ MeV and the increase in $E_{\rm MF}$ is 32.6 MeV. As a consequence the total energy is decreased by 23.9 MeV, which is 7.0 \% of the ground state energy in HF.
\begin{table}
\caption{Occupation probabilities of the proton single-particle states and the correlation energy calculated 
in TDDM for $^{40}$Ca.}
\begin{center}
\begin{tabular}{c|ccccc} \hline
  & SV &SIV &SII& SIII & SVI  \\ \hline
  $1d_{5/2}$ & 0.990&0.993&0.987 & 0.969 &0.936 \\ 
$1d_{3/2}$ &0.983&0.993&0.984  & 0.909 &0.765\\
$2s_{1/2}$ & 0.981&0.987&0.986 & 0.970 &0.944\\
$1f_{7/2}$ & 0.021&0.012&0.021 & 0.077 &0.179\\
$E_{\rm cor}$ (MeV) & -5.6&-3.6&-7.5 &-22.4&-56.5\\ \hline
\end{tabular}
\label{tab4}
\end{center}
\end{table}

\subsection{Adjustment of interaction strength}
The parameters of the Skyrme interaction have been determined to describe ground-state properties in HF which only involves relatively low relative momenta.
Therefore, it is understandable that some Skyrme parametrizations (SV, SII and SIV) induce small ground-state correlations which involve higher relative momenta.
As mentioned above, a strong cancellation occurs in the 2p--2h matrix elements of the residual interaction 
between its attractive part and repulsive momentum dependent part when the interaction has weak density dependence and strong momentum dependence.
SVI induces the strong ground-state correlations which are comparable to the results of shell-model calculations \cite{utsuno} for $^{16}$O and also
perturbation calculations for $^{40}$Ca \cite{adachi}.
Therefore, SVI may be a candidate of an effective interaction to be self-consistently used for both the mean-field part and the beyond mean-filed channels.
Since the binding energy is overestimated in TDDM, the adjustment of the force parameters is needed to reproduce the original HF energy which is close to the experimental value.
The decease in the total energy due to ground-state correlations is 10.5 MeV which is only 8.3 \% of the total energy in HF.
Therefore, a small reduction of the interaction parameters is sufficient. In fact it is found that a TDDM calculation using
2.8 \% decreased SVI parameters $t_0,~t_1,~t_2$ and $t_3$
can give the original ground-state energy in HF.  The reduction increases the HF energy by 8.0 \% 
which is compensated by the correlation energy.
In the case of SIII the reduction factor needed is 1.5 \%, which increase the HF energy by 4.3 \%.
Such a reduction factor needed is less than 1 \%
in the case of SII, SIV and SV.
The adjustment of the SVI parameters for $^{40}$Ca is also tried. In $^{40}$Ca  
the decrease in the total energy calculated with SVI is 7.0 \% of the ground state energy in HF. It is found that 
2.1 \% reduction of the SVI parameters $t_0,~t_1,~t_2$ and $t_3$ in TDDM
gives the original HF ground-state energy. The reduction increases 6.5 \% of the HF ground state energy, which compensated by $E_{\rm cor}$.

In this study only several single-particle states around the Fermi level were included. The reduction factors discussed above depend on the single-particle space used, of course.
It is worth noting that even in such small single-particle space the ground-state correlations as well as the damping properties of electric
dipole \cite{toh18,toh21}, electric quadrupole \cite{toh07,toh18} and magnetic dipole resonances \cite{toh20m} can be described to good extent.

\section{Summary}
The properties of the standard parametrizations of the Skyme force as a residual interaction 
were studied for the ground states of $^{16}$O and $^{40}$Ca
using the time-dependent density-matrix approach (TDDM) to explore the possibility that the same Skyrme interaction is consistently used for both the mean-field and the residual channels.
The three-body part ($t_3$ term) of the Skyme force was replaced by an effective density-dependent two-body interaction
which properly respect exchange properties of the residual interaction. It was found that the parametrizations with 
small $t_3$ term and large momentum-dependent $t_1$ and $t_2$ terms induce weak ground-state correlations. It was discussed that the $t_1$ and $t_2$ terms cancel the momentum independent
$t_0$ and $t_3$ terms in the 2 particle -- 2 hole matrix elements because particle states have larger momentum components than hole states. 
It was found that the Skyrme parameter set SVI which has the largest $t_3$ induces the strongest ground-state correlations comparable to the results of other theoretical calculations.
This suggests that SVI is a candidate effective interaction to be used in consistent TDDM simulations.
It was also pointed out that only a few \% reduction of the SVI parameters 
can reproduce in TDDM the original Hartree-Fock ground-state energy. 
The reduction factor mentioned above depends on the single-particle space used in the ground-state calculations.
Therefore, the approach where the same Skyrme interaction is used for both the mean-field and the residual channels may be restricted to low-energy phenomena
where a limited number of the single-particle states around the Fermi level are involved.

\let\doi\relax

\end{document}